\documentclass[%
print,
 amsmath,amssymb,
 aps
]{revtex4}

\usepackage{times}
\usepackage{graphicx}
\usepackage{subfigure}
\usepackage{dcolumn}
\usepackage{bm}
\usepackage{dsfont}
\usepackage{mathrsfs}
\usepackage[landscape,papersize={297.1mm,210mm},left=1.9cm,right=1.6cm,top=2.7cm,bottom=2.8cm]{geometry}
\usepackage[                   
            pdfstartview=FitH,
            colorlinks, 
            pdfborder=001,   
            linkcolor=blue,
            anchorcolor=blue,
            citecolor=blue,
            urlcolor=blue
            ]{hyperref}
\usepackage{amsthm}
\usepackage{pifont}
\usepackage{diagbox}
\usepackage{multirow}
\makeatletter

\newcommand{\Rmnum}[1]{\expandafter\@slowromancap\romannumeral #1@}

\makeatother

\begin{document}

\title{Quantifying multilevel coherence and multipartite correlation based on $\alpha$-affinity}

\author{Yan Hong$^{1,2}$}

\author{Mengjia Zhang$^{1}$}

\author{Limin Gao$^{1,2}$}
\email{gaoliminabc@163.com}

\author{Xianfei Qi$^{3}$}
\email{qi.xian.fei@163.com}

\author{Huaqi Zhou$^{1}$}
\email{zhouhuaqilc@163.com}

\affiliation{$^1$ School of Mathematics and Science, Hebei GEO University, Shijiazhuang 052161, China \\
$^2$ Intelligent Sensor Network Engineering Research Center of Hebei Province, Hebei GEO University, Shijiazhuang 052161, China \\
$^3$ School of Mathematics and Statistics, Shangqiu Normal University, Shangqiu 476000, China}

\begin{abstract}
Unlike standard quantum coherence, multilevel quantum coherence provides a hierarchical structure that enables a more refined characterization of quantum superposition. In this paper, we investigate multilevel coherence and introduce two $\alpha$-affinity-based indicators to quantify it, both of which satisfy several desirable properties. We further define $\alpha$-affinity-based indicators for multipartite correlation and analyze their properties. Finally, we establish relationships between these multilevel coherence indicators and the multipartite correlation indicators.
\end{abstract}

\maketitle

\section{Introduction}

Quantum coherence is an important feature of quantum mechanics and serves as a critical physical resource with significant applications in quantum algorithms \cite{HilleryPRA2016,MateraEgloff2016}, quantum thermodynamics \cite{SkrzypczykShort2014,NarasimhacharGour2015,KorzekwaLostaglioNJP2016}, and quantum metrology \cite{GiovannettiLloyd2011,TothApellanizJPA2014}.
Quantifying quantum coherence is a central topic in quantum information theory, and Baumgratz \textit{et al.} \cite{BaumgratzCramerPRL2014} established a rigorous resource-theoretic framework for quantifying quantum coherence. Based on this framework, a series of coherence measures have been proposed \cite{BaumgratzCramerPRL2014,NapoliBromleyPRL2016,YuPRA2017,StreltsovAdessoRMP2017,JinFeiPRA2018,GirolamiPRL2014,WinterYangPRL2016,RanaParasharPRA2016,RasteginPRA2016,BuSinghPRL2017,XiongKumarPRA2019,MuthuganesanPLA2021,ShiPRA2024,WangEPJD2025}, including measures based on the relative entropy of coherence, the $l_1$ norm of coherence, the robustness of coherence, and skew information.

Although significant progress has been made in quantum coherence, most studies focus on the existence of non-trivial superpositions, which is insufficient for fully understanding the role of quantum superposition. Therefore, a more refined characterization of quantum coherence is required, namely, multilevel coherence \cite{Regula2017,Chin2017,RingbauerBromleyPRX2018,RegulaPianiNJP2018,JohnstonLiPRA2018}. Ringbauer et al. generalized the standard resource-theoretic framework of coherence to multilevel coherence \cite{RingbauerBromleyPRX2018}, which includes multilevel coherence-free states, $k$-coherent-preserving operations (including $k$-incoherent operations), and a measure of multilevel coherence. Although there are various coherence measures in the standard resource theory of coherence, relatively few multilevel coherence measures have been developed within the resource theory of multilevel coherence. An efficient measure of multilevel coherence, namely the robustness of multilevel coherence $R_k^g(\rho)$, has been proposed in \cite{RingbauerBromleyPRX2018}, which satisfies several conditions for a multilevel coherence measure, such as nonnegativity, convexity, and monotonicity on average under $k$-incoherent operations. Additionally, this measure can be obtained through multilevel coherence witness experiments. The standard robustness of multilevel coherence $R_k^s(\rho)$ has also been proposed \cite{JohnstonLiPRA2018}. However, the standard robustness of multilevel coherence $R_1^s(\rho)$ is not faithful. These two robustness measures of multilevel coherence are related by $R_k^g(\rho)\leq R_k^s(\rho)$, with equality holding when $\rho$ is a pure state \cite{JohnstonLiPRA2018}.
Based on fidelity, the geometric measure of multilevel coherence $C_G^{(k)}(\rho)$ has been introduced, and an analytic expression for arbitrary pure states has been presented \cite{RegulaPianiNJP2018}.

Quantum coherence and quantum entanglement both originate from the quantum superposition principle. Numerous coherence measures \cite{BaumgratzCramerPRL2014,NapoliBromleyPRL2016,YuPRA2017,StreltsovAdessoRMP2017,JinFeiPRA2018,GirolamiPRL2014,WinterYangPRL2016,RanaParasharPRA2016,RasteginPRA2016,BuSinghPRL2017,XiongKumarPRA2019,MuthuganesanPLA2021,ShiPRA2024,WangEPJD2025} and entanglement measures \cite{VedralPlenioPRL1997,VedralPlenioPRA1998,VidalWernerPRA2002,HorodeckiRMP2009,GuhneToth2009,MaChenPRA2011,XieEberlyPRL2021,LiGaoYanPRA2024} have been proposed. It has been shown that quantum coherence and entanglement can be interconverted in bipartite and multipartite systems under certain conditions, which provides an operational connection between these two kinds of quantum resources \cite{StreltsovSinghPRL2015,MaYadinPRL2016,QiGaoYanJPA2017,ChitambarHsiehPRL2016,HuFanPRA2017,ZhuMaPRA2017,KimLeePRA2022,KodukhovKronbergPRA2023}.
Streltsov \textit{et al.} first showed that quantum coherence with respect to a fixed basis can be converted into entanglement via incoherent operations \cite{StreltsovSinghPRL2015}. Ref. \cite{MaYadinPRL2016} showed that coherence can also be converted into quantum discord. Based on the generalized Gell-Mann matrices, the coherence concurrence has been proposed in \cite{QiGaoYanJPA2017}. Moreover, its relationship with entanglement concurrence has been established, thereby providing a framework for the conversion between coherence and entanglement \cite{QiGaoYanJPA2017}. Based on fidelity, the geometric measure of $k$-coherence $C_G^{(k)}(\rho)$ and the geometric measure of $k$-partite entanglement $E_G^{(k)}(\rho)$ have been introduced \cite{RegulaPianiNJP2018}, and a coherence conversion protocol has been provided \cite{RegulaPianiNJP2018}. Fidelity and $\alpha$-affinity share many similar properties, which motivates us to study multilevel coherence from the perspective of $\alpha$-affinity.

In this paper, we further study multilevel coherence and multipartite correlations based on $\alpha$-affinity. Section II reviews the properties of $\alpha$-affinity. Section III introduces two $\alpha$-affinity-based indicators of multilevel coherence and shows that they satisfy several important properties. Section IV presents several $\alpha$-affinity-based indicators of multipartite correlations and analyzes their properties. Section V explores the relationship between the proposed multilevel coherence indicators and the multipartite correlation indicators. Finally, we summarize the results and discuss future directions in Sec. VI.

\section{$\alpha$-affinity}

In this section, we first review some basic concepts related to $\alpha$-affinity. For quantum states $\rho$ and $\sigma$ on a Hilbert space $H$, the $\alpha$-affinity is defined as $A_\alpha(\rho,\sigma)=\textrm{Tr}(\rho^\alpha\sigma^{1-\alpha})$, where $0<\alpha<1$ \cite{MuthuganesanChandrasekar2020}. When $\alpha=\frac{1}{2}$, the $\frac{1}{2}$-affinity reduces to the usual affinity \cite{LuoZhangPRA2004}. The main properties of $\alpha$-affinity are listed below.

(A1) $0\leq A_\alpha(\rho,\sigma)\leq1$ and $A_\alpha(\rho,\sigma)=1$ if and only if $\rho=\sigma$ \cite{MuthuganesanChandrasekar2020}.

(A2) The $\alpha$-affinity is invariant under unitary transformations: $A_\alpha(U\rho U^\dagger,U\sigma U^\dagger)=A_\alpha(\rho,\sigma)$ for any unitary operator $U$.

(A3) The $\alpha$-affinity is multiplicative: $A_\alpha(\rho_1\otimes\rho_2,\sigma_1\otimes\sigma_2)=A_\alpha(\rho_1,\sigma_1)A_\alpha(\rho_2,\sigma_2)$.

(A4) The $\alpha$-affinity is jointly concave, i.e., $A_\alpha(\sum\limits_i\lambda_i\rho_i,\sum\limits_i\lambda_i\sigma_i)\geq \sum\limits_i\lambda_iA_\alpha(\rho_i,\sigma_i)$, where $\lambda_i\geq0$, $\sum\limits_i\lambda_i=1$,  and both $\rho_i$ and $\sigma_i$ are quantum states on a Hilbert space $H$ \cite{MuthuganesanChandrasekar2020}.

(A5) The $\alpha$-affinity is monotonic under a CPTP map $\Lambda$: $A_\alpha(\rho,\sigma)\leq A_\alpha(\Lambda(\rho),\Lambda(\sigma))$ \cite{MuthuganesanChandrasekar2020}.

(A6) For a CPTP map with Kraus operators $\{K_i\}$, let $p_i=\textrm{Tr}(K_i\rho K_i^\dagger), q_i=\textrm{Tr}(K_i\sigma K_i^\dagger), \rho_i=\dfrac{K_i\rho K_i^\dagger}{p_i}$, and $\sigma_i=\dfrac{K_i\sigma K_i^\dagger}{q_i}$. Then the following inequality holds:
\begin{equation}\label{PropositionB5}
\begin{aligned}
\sum\limits_ip_i\left\{1-\Big[A_\alpha(\rho_i,\sigma_i)\Big]^{\frac{1}{\alpha}}\right\}\leq1-\Big[A_\alpha(\rho,\sigma)\Big]^{\frac{1}{\alpha}}.
\end{aligned}
\end{equation}

The proofs of items (A2) and (A3) are given in Appendix A. The proof of item (A6) is given in Appendix B.

\section{Quantification of multilevel coherence}

A rigorous resource-theoretic framework for quantifying quantum coherence has been introduced, encompassing definitions of incoherent states, incoherent operations, and coherence measures \cite{BaumgratzCramerPRL2014}.
For a Hilbert space $H$ with $\dim(H)=d$, let $\{|i\rangle: i=0,1,\cdots,d-1\}$ be an orthonormal basis of $H$.
A quantum state is called incoherent if its density matrix is diagonal with respect to a chosen orthonormal basis $\{|i\rangle: i=0,1,\cdots,d-1\}$; otherwise, it is coherent. The set of all incoherent states is denoted by $I$.
Any incoherent state $\rho\in I$ admits a representation of the form
$\rho=\sum\limits_{i=0}^{d-1}p_i|i\rangle\langle i|$.
A completely positive trace-preserving (CPTP) map $\Lambda$, with Kraus decomposition $\Lambda(\rho)=\sum\limits_{n}K_n\rho K_n^\dagger$, is called an incoherent operation if each Kraus operator maps incoherent states to incoherent states,
 i.e., $K_n\rho K_n^\dagger/\textrm{Tr}(K_n\rho K_n^\dagger)\in I$ for any $\rho\in I$.

Based on the concepts of incoherent states and incoherent operations, a framework for quantifying quantum coherence has been established. Within this framework, a coherence measure $C$ is defined as a mapping from quantum states to nonnegative real numbers that satisfies the following conditions \cite{BaumgratzCramerPRL2014}: (1) Faithfulness; that is,  the measure is nonnegative for all states and vanishes if and only if the state is incoherent.
(2) Monotonicity; that is,  $C(\Lambda(\rho))\leq C(\rho)$ for any quantum state $\rho$ and incoherent operation $\Lambda$.
(3) Strong monotonicity; that is, $\sum\limits_{i}p_iC(\rho_i)\leq C(\rho)$ with $p_i=\textrm{Tr}(K_i\rho K_i^\dagger), \rho_i=\dfrac{K_i\rho K_i^\dagger}{\textrm{Tr}(K_i\rho K_i^\dagger)}$, and $\{K_i\}$ being incoherent Kraus operators.
(4) Convexity; that is, $C(\sum\limits_ip_i\rho_i )\leq\sum\limits_i p_i C(\rho_i)$ for any  probability distribution $\{p_i\}$ and any set of states $\{\rho_i\}$. Additionally, strong monotonicity and convexity imply monotonicity.

For a Hilbert space $H$, let $\{|i\rangle: i=0,1,\cdots,d-1\}$ be an orthonormal basis of $H$.
If a pure state $|\psi\rangle$ can be written as $|\psi\rangle=\sum\limits_{i=0}^{d-1}c_i|i\rangle$ with $\sum\limits_{i=0}^{d-1}|c_i|^2=1$, then
the number of non-zero coefficients $c_i$ is called the coherence rank of a pure state $|\psi\rangle$, denoted by $R_c(|\psi\rangle)$ \cite{RingbauerBromleyPRX2018,RegulaPianiNJP2018}.
For a mixed state $\rho$, its coherence number is defined as $R_c(\rho)=\min\limits_{\{p_i,|\psi_i\rangle\}}\max\limits_iR_c(|\psi_i\rangle)$,
where the minimum is taken over all pure state decompositions $\{p_i,|\psi_i\rangle\}$ of $\rho$ \cite{RegulaPianiNJP2018}.

Let the set $I_k$ consist of convex combinations of pure states with coherence rank at most
$k$; that is,
\begin{equation*}
\begin{aligned}
I_k:=\Big\{\sum\limits_ip_i|\psi_i\rangle\langle\psi_i|: p_i\geq0, \sum\limits_ip_i=1, R_c(|\psi_i\rangle)\leq k\Big\},
\end{aligned}
\end{equation*}
where $1\leq k\leq d$ \cite{RingbauerBromleyPRX2018}. $I_d$ is the set of all quantum states. One readily verifies that $I_1\subset I_2\subset\cdots\subset I_d$. Moreover, $I_k$ is the set of all $(k+1)$-level coherence-free states \cite{RingbauerBromleyPRX2018}.

Let $\Lambda$ be a completely positive trace-preserving (CPTP) map.
The CPTP map $\Lambda$ is called a $k$-coherent-preserving operation if $\Lambda(\rho)\in I_k$ for any $(k+1)$-level coherence-free state $\rho$,
i.e., $\Lambda(I_k)\subseteq I_k$  \cite{RingbauerBromleyPRX2018}.
For any $(k+1)$-level coherence-free state $\rho$, suppose that there exists a Kraus decomposition $\{K_i\}$ of the CPTP map $\Lambda$ such that  $\dfrac{K_i\rho K_i^\dagger}{\textrm{Tr}(K_i\rho K_i^\dagger)}\in I_k$ for every $i$. Then the map $\Lambda$ is called a $k$-incoherent operation \cite{RingbauerBromleyPRX2018}.
The $k$-incoherent operations are a subset of $k$-coherent-preserving operations \cite{RingbauerBromleyPRX2018}.

From the above concepts, one can derive the relationship between quantum coherence and multilevel quantum coherence. For example, the set of all $2$-level coherence-free states $I_1$ is the set of incoherent states $I$, namely, $I_1=I$, and a $1$-incoherent operation is an incoherent operation.

For a quantum state $\rho$ on a Hilbert space $H$ with $\dim(H)=d$, we can define two indicators for characterizing multilevel coherence,
\begin{equation*}
\begin{aligned}
\mathcal{C}_\alpha^{k}(\rho):=\min\limits_{\sigma\in I_{k-1}}\Big[1-A_\alpha(\rho,\sigma)\Big]=1-\max\limits_{\sigma\in I_{k-1}}A_\alpha(\rho,\sigma),
\end{aligned}
\end{equation*}
\begin{equation*}
\begin{aligned}
\mathfrak{C}_\alpha^{k}(\rho):=\min\limits_{\sigma\in I_{k-1}}\left\{1-\Big[A_\alpha(\rho,\sigma)\Big]^{\frac{1}{\alpha}}\right\}=1-\max\limits_{\sigma\in I_{k-1}}\Big[A_\alpha(\rho,\sigma)\Big]^{\frac{1}{\alpha}}.
\end{aligned}
\end{equation*}

\emph{Theorem 1.} The indicators $\mathcal{C}_\alpha^{k}(\rho)$ and $\mathfrak{C}_\alpha^{k}(\rho)$ satisfy the following properties.

(P$'$1) $\mathcal{C}_\alpha^{k}(\rho)\geq0$ and $\mathfrak{C}_\alpha^{k}(\rho)\geq0$ for any quantum state $\rho$, $\mathcal{C}_\alpha^{k}(\rho)=0$ and $\mathfrak{C}_\alpha^{k}(\rho)=0$ if and only if $\rho\in I_{k-1}$.

(P$'$2) $\mathcal{C}_\alpha^{k}(\rho)$ is convex; that is,
\begin{equation*}
\begin{aligned}
\mathcal{C}_\alpha^{k}(\sum\limits_ip_i\rho_i )\leq&\sum\limits_i p_i \mathcal{C}_\alpha^{k}(\rho_i),
\end{aligned}
\end{equation*}
where $p_i\geq0$ and $\sum\limits_ip_i=1$.

(P$'$3) $\mathfrak{C}_{\alpha}^{k}(\rho)$ is nonincreasing on average under a $(k-1)$-incoherent operation with Kraus operators $\{K_i\}$,
\begin{equation*}
\begin{aligned}
\mathfrak{C}_{\alpha}^{k}(\rho)\geq&\sum\limits_ip_i\mathfrak{C}_{\alpha}^{k}(\rho_i),
\end{aligned}
\end{equation*}
 where $p_i=\textrm{Tr}(K_i\rho K_i^\dagger)$ and $\rho_i=\dfrac{K_i\rho K_i^\dagger}{p_i}$.

(P$'$4) $\mathcal{C}_\alpha^{k}(\rho)$ and  $\mathfrak{C}_\alpha^{k}(\rho)$ are both nonincreasing under a $(k-1)$-coherence-preserving operation $\Lambda$,
\begin{equation*}
\begin{aligned}
\mathcal{C}_\alpha^{k}(\rho)\geq&\mathcal{C}_\alpha^{k}(\Lambda(\rho)),\\
\mathfrak{C}_\alpha^{k}(\rho)\geq&\mathfrak{C}_\alpha^{k}(\Lambda(\rho)).
\end{aligned}
\end{equation*}

(P$'$5) $\mathcal{C}_\alpha^{k}(\rho)$ and $\mathfrak{C}_\alpha^{k}(\rho)$ satisfy the following inequalities:
\begin{equation*}
\begin{aligned}
\mathcal{C}_\alpha^{(k-1)^m+1}(\bigotimes\limits_{j=1}^m\rho_j )\leq&\sum\limits_{j=1}^m\mathcal{C}_\alpha^{k}(\rho_j),\\
\mathfrak{C}_\alpha^{(k-1)^m+1}(\bigotimes\limits_{j=1}^m\rho_j )\leq&\sum\limits_{j=1}^m\mathfrak{C}_\alpha^{k}(\rho_j),
\end{aligned}
\end{equation*}
where each $\rho_j$ is a quantum state on $H$.

\emph{Proof}:
(P$'$1) Since $0\leq A_\alpha(\rho,\sigma)\leq1$, we have $0\leq \mathcal{C}_\alpha^{k}(\rho)\leq1$ and $0\leq \mathfrak{C}_\alpha^k(\rho)\leq1$.
Since $A_\alpha(\rho,\sigma)=1$ if and only if $\rho=\sigma$, the definitions of $\mathcal{C}_\alpha^{k}(\rho)$ and $\mathfrak{C}_\alpha^{k}(\rho)$ imply that $\mathcal{C}_\alpha^{k}(\rho)=0$ and $\mathfrak{C}_\alpha^{k}(\rho)=0$ if and only if $\rho\in I_{k-1}$.

(P$'$2) Let $\rho=\sum\limits_i p_i\rho_i$, where $p_i\geq0$ and $\sum\limits_i p_i=1$. Denote by $\sigma_i^*$ the state in $I_{k-1}$ minimizing
$1-A_\alpha(\rho_i,\sigma)$ over $\sigma\in I_{k-1}$, so that $\mathcal{C}_\alpha^{k}(\rho_i)=1-A_\alpha(\rho_i,\sigma^*_i)$. Then
\begin{align}
\mathcal{C}_\alpha^{k}(\sum\limits_ip_i\rho_i )
\leq&1-A_\alpha(\sum\limits_ip_i\rho_i,\sum\limits_ip_i\sigma^*_i)\label{PP21}\\
\leq&1-\sum\limits_ip_iA_\alpha(\rho_i,\sigma^*_i)\label{PP22}\\
=&\sum\limits_ip_i\Big[1-A_\alpha(\rho_i,\sigma^*_i)\Big]\label{}\nonumber\\
=&\sum\limits_ip_i\mathcal{C}_\alpha^{k}(\rho_i ).\label{}\nonumber
\end{align}
Here, inequality (\ref{PP21}) holds because  $\sum\limits_ip_i\sigma^*_i\in I_{k-1}$ when $\sigma^*_i\in I_{k-1}$.
Inequality (\ref{PP22}) holds by the joint concavity of $\alpha$-affinity.

(P$'$3) Let $\sigma^*\in I_{k-1}$ be the minimizer of $1-\Big[A_{\alpha}(\rho,\sigma)\Big]^{\frac{1}{\alpha}}$ over $\sigma\in I_{k-1}$. Then $\mathfrak{C}_{\alpha}^{k}(\rho)=1-\Big[A_{\alpha}(\rho,\sigma^*)\Big]^{\frac{1}{\alpha}}$. Next, we obtain
\begin{align}
\mathfrak{C}_{\alpha}^{k}(\rho)=&1-\Big[A_\alpha(\rho,\sigma^*)\Big]^{\frac{1}{\alpha}}\label{}\nonumber\\
\geq&\sum\limits_ip_i\left\{1-\Big[A_{\alpha}(\rho_i,\dfrac{K_i\sigma^* K_i^\dagger}{\textrm{Tr}(K_i\sigma^* K_i^\dagger)})\Big]^{\frac{1}{\alpha}}\right\}\label{PP32}\\
\geq&\sum\limits_ip_i\mathfrak{C}_{\alpha}^{k}(\rho_i)\label{PP33}.
 \end{align}
Here, inequality  (\ref{PP32}) follows from inequality (\ref{PropositionB5}). Inequality (\ref{PP33}) holds because
$\dfrac{K_i\sigma^* K_i^\dagger}{\textrm{Tr}(K_i\sigma^* K_i^\dagger)}\in I_{k-1}$ when $\sigma^*\in I_{k-1}$
for a $(k-1)$-incoherent operation with Kraus operators $\{K_i\}$.

(P$'$4) For any $(k-1)$-coherence-preserving operation $\Lambda$, we have
\begin{align}
\mathfrak{C}_\alpha^{k}(\Lambda(\rho))=&\min\limits_{\sigma\in I_{k-1}}\left\{1-\Big[A_\alpha(\Lambda(\rho),\sigma)\Big]^{\frac{1}{\alpha}}\right\}\label{}\nonumber\\
\leq&\min\limits_{\sigma\in I_{k-1}}\left\{1-\Big[A_\alpha(\Lambda(\rho),\Lambda(\sigma))\Big]^{\frac{1}{\alpha}}\right\}\label{PP403}\\
\leq&\min\limits_{\sigma\in I_{k-1}}\left\{1-\Big[A_\alpha(\rho,\sigma)\Big]^{\frac{1}{\alpha}}\right\}\label{PP404}\\
=&\mathfrak{C}_\alpha^{k}(\rho).\label{}\nonumber
 \end{align}
Inequality  (\ref{PP403}) holds because $\Lambda(\sigma) \in I_{k-1}$ when $\sigma\in I_{k-1}$
under a $(k-1)$-coherence-preserving operation $\Lambda$, and inequality (\ref{PP404}) is a direct application of the monotonicity property (A5). An analogous argument shows that $\mathcal{C}_\alpha^{k}(\rho)$ is also nonincreasing under a $(k-1)$-coherence-preserving operation $\Lambda$.

(P$'$5) For a quantum state $\rho_j$ on $H$ with $1\leq j\leq m$, let $\sigma_j^{*}\in I_{k-1}$ be the maximizer of $A_\alpha(\rho_j,\sigma_j)$ over $\sigma_j\in I_{k-1}$. Then $\mathfrak{C}_\alpha^{k}(\rho_j)=1-\Big[A_\alpha(\rho_j,\sigma_j^{*})\Big]^{\frac{1}{\alpha}}$.
Thus we have
\begin{align}
\mathfrak{C}_\alpha^{(k-1)^m+1}(\bigotimes\limits_{j=1}^m\rho_j )
\leq&1-\Big[A_\alpha(\bigotimes\limits_{j=1}^m\rho_j,\bigotimes\limits_{j=1}^m\sigma^*_j)\Big]^{\frac{1}{\alpha}}\label{PP511}\\
=&1-\prod\limits_{j=1}^m\Big[A_\alpha(\rho_j,\sigma^*_j)\Big]^{\frac{1}{\alpha}}\label{PP521}\\
\leq&\sum\limits_{j=1}^m\left\{1-\Big[A_\alpha(\rho_j,\sigma^*_j)\Big]^{\frac{1}{\alpha}}\right\}\label{PP531}\\
=&\sum\limits_{j=1}^m\mathfrak{C}_\alpha^{k}(\rho_j ).\label{}\nonumber
\end{align}
Here, inequality (\ref{PP511}) holds because $\bigotimes\limits_{j=1}^m\sigma_j^*\in I_{(k-1)^m}$ when $\sigma_j^*\in I_{k-1}$.
Equation (\ref{PP521}) follows from property (A3) of $\alpha$-affinity, and inequality (\ref{PP531}) follows from inequality $1-\prod\limits_{i=1}^mx_i\leq\sum\limits_{i=1}^m(1-x_i)$ with $0\leq x_i\leq 1$.
We can similarly derive that $\mathcal{C}_\alpha^{k}(\rho)$ also satisfies this property.

\section{Quantification of multipartite correlation}

For an $N$-partite pure state $|\psi\rangle\in H_1\otimes H_2\otimes \cdots \otimes H_N$,  if there exists a partition of the form $\gamma := \gamma_1|\cdots|\gamma_{|\gamma|}$ such that $|\psi\rangle=|\psi_{\gamma_1}\rangle|\psi_{\gamma_2}\rangle\cdots|\psi_{\gamma_{|\gamma|}}\rangle$, then $|\psi\rangle$ is said to be $k$-separable if $|\gamma|=k$, and $k$-producible if $n_{\gamma_i}\leq k$ for $1\leq i\leq |\gamma|$ \cite{GuhneToth2009}.
If an $N$-partite mixed state $\rho$ can be represented as a convex combination of $k$-separable (respectively, $k$-producible) pure states, it is said to be $k$-separable (respectively, $k$-producible) \cite{GuhneToth2009}. A quantum state $\rho$ that is not $k$-separable is said to be $k$-nonseparable; similarly, a state that is not $k$-producible is said to contain $(k+1)$-partite entanglement. Let $S_k$ be the set of all $k$-separable states, and let $P_k$ be the set of all $k$-producible states. Then
$S_1\supset S_2\supset\cdots \supset S_N$,  $P_1\subset P_2\subset\cdots \subset P_N$. Both $S_1$ and $P_N$ coincide with the set of all quantum states. If a quantum state is $k$-separable but not $(k+1)$-separable, we say that its separability depth is $k$, denoted by $D_S(\rho)=k$. If a quantum state is $k$-producible but not $(k-1)$-producible, we call it genuinely $k$-partite entangled, and  we say its entanglement depth is $k$, denoted by $D_E(\rho)=k$. In particular, an $N$-separable state (or a $1$-producible state) coincides with the conventional notion of a fully separable state; that is, $D_S(\rho)=N$ (or  $D_E(\rho)=1$) if and only if the quantum state $\rho$ is fully separable.

For an $N$-partite quantum state $\rho\in H_1\otimes H_2\otimes \cdots \otimes H_N$, we define several multipartite correlation indicators as follows:
\begin{equation*}
\begin{aligned}
\mathcal{S}_\alpha^k(\rho):=&\min\limits_{\sigma\in S_k}\Big[1-A_\alpha(\rho,\sigma)\Big]=1-\max\limits_{\sigma\in S_k}A_\alpha(\rho,\sigma),
\end{aligned}
\end{equation*}
\begin{equation*}
\begin{aligned}
\mathfrak{S}_\alpha^k(\rho):=&\min\limits_{\sigma\in S_k}\left\{1-\Big[A_\alpha(\rho,\sigma)\Big]^{\frac{1}{\alpha}}\right\}=1-\max\limits_{\sigma\in S_k}\Big[A_\alpha(\rho,\sigma)\Big]^{\frac{1}{\alpha}},
\end{aligned}
\end{equation*}
 \begin{equation*}
\begin{aligned}
\widetilde{E}_\alpha^{k}(\rho):=\min\limits_{\sigma\in P_{k-1}}\Big[1-A_\alpha(\rho,\sigma)\Big]=1-\max\limits_{\sigma\in P_{k-1}}A_\alpha(\rho,\sigma),
\end{aligned}
\end{equation*}
 \begin{equation*}
\begin{aligned}
\widetilde{\mathcal{E}}_\alpha^{k}(\rho):=\min\limits_{\sigma\in P_{k-1}}\left[1-\Big[A_\alpha(\rho,\sigma)\Big]^{\frac{1}{\alpha}}\right\}=1-\max\limits_{\sigma\in P_{k-1}}\Big[A_\alpha(\rho,\sigma)\Big]^{\frac{1}{\alpha}}.
\end{aligned}
\end{equation*}

\emph{Theorem 2.} $\mathcal{S}_\alpha^k(\rho), \mathfrak{S}_\alpha^k(\rho), \widetilde{E}_\alpha^{k}(\rho)$, and $\widetilde{\mathcal{E}}_\alpha^{k}(\rho)$ satisfy the following properties.

(P1) $\mathcal{S}_\alpha^k(\rho)\geq0$ and $\mathfrak{S}_\alpha^k(\rho)\geq0$ for any quantum state $\rho$,  $\mathcal{S}_\alpha^k(\rho)=0$ and $\mathfrak{S}_\alpha^k(\rho)=0$ if and only if $\rho\in S_k$.
$\widetilde{E}_\alpha^{k}(\rho)\geq0$ and $\widetilde{\mathcal{E}}_\alpha^{k}(\rho)\geq0$ for any quantum state $\rho$, $\widetilde{E}_\alpha^{k}(\rho)=0$  and $\widetilde{\mathcal{E}}_\alpha^{k}(\rho)=0$ if and only if $\rho \in P_{k-1}$.

(P2) $\mathcal{S}_\alpha^k(\rho), \mathfrak{S}_\alpha^k(\rho), \widetilde{E}_\alpha^{k}(\rho)$, and $\widetilde{\mathcal{E}}_\alpha^{k}(\rho)$ are invariant under local unitary transformations:
\begin{equation*}
\begin{aligned}
\mathcal{S}_\alpha^k\big((U_1\otimes U_2\otimes\cdots\otimes U_N)\rho
(U_1^\dagger\otimes U_2^\dagger\otimes\cdots\otimes U_N^\dagger)\big)=&\mathcal{S}_\alpha^k(\rho),\\
\mathfrak{S}_\alpha^k\big((U_1\otimes U_2\otimes\cdots\otimes U_N)\rho
(U_1^\dagger\otimes U_2^\dagger\otimes\cdots\otimes U_N^\dagger)\big)=&\mathfrak{S}_\alpha^k(\rho),\\
\widetilde{E}_\alpha^{k}\big((U_1\otimes U_2\otimes\cdots\otimes U_N)\rho
(U_1^\dagger\otimes U_2^\dagger\otimes\cdots\otimes U_N^\dagger)\big)=&\widetilde{E}_\alpha^{k}(\rho),\\
\widetilde{\mathcal{E}}_\alpha^{k}\big((U_1\otimes U_2\otimes\cdots\otimes U_N)\rho
(U_1^\dagger\otimes U_2^\dagger\otimes\cdots\otimes U_N^\dagger)\big)=&\widetilde{\mathcal{E}}_\alpha^{k}(\rho).
\end{aligned}
\end{equation*}

(P3) $\mathcal{S}_\alpha^k(\rho)$ and $\widetilde{E}_\alpha^{k}(\rho)$ are both convex; that is,
\begin{equation*}
\begin{aligned}
\mathcal{S}_\alpha^k(\sum\limits_ip_i\rho_i )\leq&\sum\limits_i p_i \mathcal{S}_\alpha^k(\rho_i),\\
\widetilde{E}_\alpha^{k}(\sum\limits_ip_i\rho_i )\leq&\sum\limits_i p_i\widetilde{E}_\alpha^{k}(\rho_i),
\end{aligned}
\end{equation*}
where $p_i\geq0$ and $\sum\limits_ip_i=1.$

(P4) For a quantum state $\rho$ on
$H_1 \otimes H_2 \otimes \cdots \otimes H_N$,
both $\mathfrak{S}_{\alpha}^k(\rho)$ and
$\widetilde{\mathcal{E}}_{\alpha}^{k}(\rho)$
are nonincreasing on average under separable operations with product Kraus operators
$\{K_i:=\bigotimes_{j=1}^N \Gamma_i^{(j)}\}$,
\begin{equation*}
\begin{aligned}
\mathfrak{S}_{\alpha}^k(\rho)\geq&
\sum\limits_ip_i\mathfrak{S}_{\alpha}^k(\rho_i),\\
\widetilde{\mathcal{E}}_{\alpha}^{k}(\rho)\geq&
\sum\limits_ip_i\widetilde{\mathcal{E}}_{\alpha}^{k}(\rho_i),
\end{aligned}
\end{equation*}
where
$p_i=\textrm{Tr}(K_i\rho K_i^\dagger)$ and
$\rho_i=\dfrac{K_i\rho K_i^\dagger}{p_i}$.

(P5) $\mathcal{S}_\alpha^k(\rho)$,
$\mathfrak{S}_\alpha^k(\rho)$,
$\widetilde{E}_\alpha^{k}(\rho)$, and
$\widetilde{\mathcal{E}}_\alpha^{k}(\rho)$
are nonincreasing under separable operations $\Lambda$,
\begin{equation*}
\begin{aligned}
\mathcal{S}_\alpha^k(\rho )\geq&
\mathcal{S}_\alpha^k(\Lambda(\rho)),\\
\mathfrak{S}_\alpha^k(\rho )\geq&
\mathfrak{S}_\alpha^k(\Lambda(\rho)),\\
\widetilde{E}_\alpha^{k}(\rho )\geq&
\widetilde{E}_\alpha^{k}(\Lambda(\rho)),\\
\widetilde{\mathcal{E}}_\alpha^{k}(\rho )\geq&
\widetilde{\mathcal{E}}_\alpha^{k}(\Lambda(\rho)).
\end{aligned}
\end{equation*}

(P6) $\mathcal{S}_\alpha^k(\rho), \mathfrak{S}_\alpha^k(\rho), \widetilde{E}_\alpha^{k}(\rho)$, and $\widetilde{\mathcal{E}}_\alpha^{k}(\rho)$ satisfy subadditivity,
\begin{equation*}
\begin{aligned}
\mathcal{S}_\alpha^k(\bigotimes\limits_{j=1}^m\rho_j )\leq&\sum\limits_{j=1}^m\mathcal{S}_\alpha^k(\rho_j),\\
\mathfrak{S}_\alpha^k(\bigotimes\limits_{j=1}^m\rho_j )\leq&\sum\limits_{j=1}^m\mathfrak{S}_\alpha^k(\rho_j),\\
\widetilde{E}_\alpha^{k}(\bigotimes\limits_{j=1}^m\rho_j )\leq&\sum\limits_{j=1}^m\widetilde{E}_\alpha^{k}(\rho_j),\\
\widetilde{\mathcal{E}}_\alpha^{k}(\bigotimes\limits_{j=1}^m\rho_j )\leq&\sum\limits_{j=1}^m\widetilde{\mathcal{E}}_\alpha^{k}(\rho_j),
\end{aligned}
\end{equation*}
where each $\rho_j$ is an $N$-partite quantum state on $H_1\otimes H_2\otimes\cdots\otimes H_N$.

\emph{Proof}:
(P1) Since $0\leq A_\alpha(\rho,\sigma)\leq1$,  we have $0\leq \mathcal{S}_\alpha^k(\rho)\leq1$, $0\leq \mathfrak{S}_\alpha^k(\rho)\leq1$, $0\leq\widetilde{E}_\alpha^{k}(\rho)\leq1$, and $0\leq\widetilde{\mathcal{E}}_\alpha^{k}(\rho)\leq1$.
Since $A_\alpha(\rho,\sigma)=1$ if and only if $\rho=\sigma$, the definitions of  $\mathcal{S}_\alpha^k(\rho), \mathfrak{S}_\alpha^k(\rho), \widetilde{E}_\alpha^{k}(\rho)$, and $\widetilde{\mathcal{E}}_\alpha^{k}(\rho)$ imply that $\mathcal{S}_\alpha^k(\rho)=\mathfrak{S}_\alpha^k(\rho)=0$ if and only if $\rho\in S_{k}$,  and $\widetilde{E}_\alpha^{k}(\rho)=\widetilde{\mathcal{E}}_\alpha^{k}(\rho)=0$ if and only if $\rho\in P_{k-1}$.

(P2) For a quantum state $\rho$ on $H_1\otimes H_2\otimes\cdots\otimes H_N$, let $U_i$ be a unitary transformation acting on subsystem $H_i$. Then
\begin{align}
&\mathcal{S}_\alpha^k\big((U_1\otimes U_2\otimes\cdots\otimes U_N)\rho
(U_1^\dagger\otimes U_2^\dagger\otimes\cdots\otimes U_N^\dagger)\big)\label{}\nonumber\\
=&\min\limits_{\sigma\in S_k}
\Big[1-A_\alpha\big((U_1\otimes U_2\otimes\cdots\otimes U_N)\rho
(U_1^\dagger\otimes U_2^\dagger\otimes\cdots\otimes U_N^\dagger),\sigma\big)\Big]\label{}\nonumber\\
=&\min\limits_{\sigma\in S_k}
\Big[1-A_\alpha\Big((U_1\otimes U_2\otimes\cdots\otimes U_N)\rho
(U_1^\dagger\otimes U_2^\dagger\otimes\cdots\otimes U_N^\dagger),
(U_1\otimes U_2\otimes\cdots\otimes U_N)\sigma
(U_1^\dagger\otimes U_2^\dagger\otimes\cdots\otimes U_N^\dagger)\Big)\Big]\label{P21}\\
=&\min\limits_{\sigma\in S_k}\Big[1-A_\alpha(\rho,\sigma)\Big]\label{P22}\\
=&\mathcal{S}_\alpha^k(\rho).\label{}\nonumber
\end{align}
Equation (\ref{P21}) holds because $(U_1\otimes U_2\otimes\cdots\otimes U_N)
\sigma (U_1^\dagger\otimes U_2^\dagger\otimes\cdots\otimes U_N^\dagger)$ is $k$-separable if and only if $\sigma$ is $k$-separable.
Equation (\ref{P22}) follows from the unitary invariance of $\alpha$-affinity.
Similarly, we can prove that $\mathfrak{S}_\alpha^k(\rho), \widetilde{E}_\alpha^{k}(\rho)$, and $\widetilde{\mathcal{E}}_\alpha^{k}(\rho)$ remain invariant under local unitary transformations.

(P3) Let $\rho=\sum\limits_i p_i\rho_i$, where $p_i\geq0$ and $\sum\limits_ip_i=1$. Denote by $\sigma_i^*$ the state in $S_k$ minimizing
$1-A_{\alpha}(\rho_i, \sigma)$ over $\sigma\in S_k$, so that $\mathcal{S}_{\alpha}^k(\rho_i) = 1 - A_{\alpha}(\rho_i, \sigma_i^*)$. Then
\begin{align}
\mathcal{S}_\alpha^k(\sum\limits_ip_i\rho_i )
\leq&1-A_\alpha(\sum\limits_ip_i\rho_i,\sum\limits_ip_i\sigma^*_i)\label{P31}\\
\leq&1-\sum\limits_ip_iA_\alpha(\rho_i,\sigma^*_i)\label{P32}\\
=&\sum\limits_ip_i\Big[1-A_\alpha(\rho_i,\sigma^*_i)\Big]\label{}\nonumber\\
=&\sum\limits_ip_i\mathcal{S}_\alpha^k(\rho_i ).\label{}\nonumber
\end{align}
Here, inequality (\ref{P31}) holds because $\sum\limits_ip_i\sigma^*_i\in S_k$ if $\sigma^*_i\in S_k$.
Inequality (\ref{P32}) holds by the joint concavity of $\alpha$-affinity.
Convexity of  $\widetilde{E}_\alpha^{k}(\rho)$ follows by an identical argument.

(P4) Let $\sigma^{*} \in S_k$ be the minimizer of $1-\Big[A_\alpha(\rho, \sigma)\Big]^{\frac{1}{\alpha}}$ over $\sigma\in S_k$. Then $\mathfrak{S}_{\alpha}^k(\rho) = 1 - \Big[A_\alpha(\rho, \sigma^*)\Big]^{\frac{1}{\alpha}}$. Then
\begin{align}
\mathfrak{S}_{\alpha}^k(\rho )=&1-\Big[A_{\alpha}(\rho,\sigma^*)\Big]^{\frac{1}{\alpha}}\label{}\nonumber\\
\geq&\sum\limits_ip_i\left\{1-\Big[A_{\alpha}(\rho_i,\dfrac{K_i\sigma^* K_i^\dagger}{\textrm{Tr}(K_i\sigma^* K_i^\dagger)})\Big]^{\frac{1}{\alpha}}\right\}\label{P41}\\
\geq&\sum\limits_i p_i\mathfrak{S}_{\alpha}^k(\rho_i ).\label{P42}
\end{align}
Inequality (\ref{P41}) follows from (\ref{PropositionB5}).
Inequality (\ref{P42}) holds because $\dfrac{K_i\sigma^* K_i^\dagger}{\textrm{Tr}(K_i\sigma^* K_i^\dagger)}\in S_k$ when $\sigma^*\in S_k $ with Kraus operators $K_i=\bigotimes\limits_{j=1}^N\Gamma_i^{(j)}$.
The proof that $\widetilde{\mathcal{E}}_\alpha^{k}(\rho)$ is also nonincreasing on average under separable operations with product Kraus operators $\{K_i:=\bigotimes_{j=1}^N\Gamma_i^{(j)}\}$ is analogous.

(P5) For any separable operation $\Lambda$, we obtain
\begin{align}
\mathcal{S}_\alpha^{k}[\Lambda(\rho)]=&\min\limits_{\sigma\in S_{k}}\Big[1-A_\alpha(\Lambda(\rho),\sigma)\Big]\label{}\nonumber\\
\leq&\min\limits_{\sigma\in S_{k}}\Big[1-A_\alpha(\Lambda(\rho),\Lambda(\sigma))\Big]\label{PP501}\\
\leq&\min\limits_{\sigma\in S_{k}}\Big[1-A_\alpha(\rho,\sigma)\Big]\label{PP502}\\
=&\mathcal{S}_\alpha^{k}(\rho).\label{}\nonumber
 \end{align}
Inequalities (\ref{PP501}) and (\ref{PP502}) hold because $\Lambda(\sigma)\in S_k$ for any $\sigma\in S_k$ and because of property (A5) of $\alpha$-affinity, respectively. Similarly, we can prove that $\mathfrak{S}_\alpha^k(\rho), \widetilde{E}_\alpha^{k}(\rho)$, and $\widetilde{\mathcal{E}}_\alpha^{k}(\rho)$  are nonincreasing under separable operations $\Lambda$.

(P6) For a quantum state $\rho_j$ on $H_1\otimes H_2\otimes\cdots\otimes H_N$ with $1\leq j\leq m$, let $\sigma_j^* \in S_k$ and $\widetilde{\sigma}_j^* \in P_{k-1}$ be the maximizers of $A_\alpha(\rho_j, \sigma_j)$ over $\sigma_j\in S_k$ and $\sigma_j\in P_{k-1}$, respectively. Thus we have
\begin{align}
\mathcal{S}_\alpha^k(\bigotimes\limits_{j=1}^m\rho_j )
\leq&\mathcal{S}_\alpha^{mk}(\bigotimes\limits_{j=1}^m\rho_j )\label{P61}\\
\leq&1-A_\alpha(\bigotimes\limits_{j=1}^m\rho_j,\bigotimes\limits_{j=1}^m\sigma^*_j)\label{P62}\\
=&1-\prod\limits_{j=1}^mA_\alpha(\rho_j,\sigma^*_j)\label{P63}\\
\leq&\sum\limits_{j=1}^m\Big[1-A_\alpha(\rho_j,\sigma^*_j)\Big]\label{P64}\\
=&\sum\limits_{j=1}^m\mathcal{S}_\alpha^k(\rho_j ),\label{}\nonumber
\end{align}
\begin{align}
\widetilde{E}_\alpha^{k}(\bigotimes\limits_{j=1}^m\rho_j )
\leq&1-A_\alpha(\bigotimes\limits_{j=1}^m\rho_j,\bigotimes\limits_{j=1}^m\widetilde{\sigma}^*_j)\label{P65}\\
=&1-\prod\limits_{j=1}^mA_\alpha(\rho_j,\widetilde{\sigma}^*_j)\label{P66}\\
\leq&\sum\limits_{j=1}^m\Big[1-A_\alpha(\rho_j,\widetilde{\sigma}^*_j)\Big]\label{P67}\\
=&\sum\limits_{j=1}^m\widetilde{E}_\alpha^{k}(\rho_j ).\label{}\nonumber
\end{align}
Here, inequality (\ref{P61}) holds because $\mathcal{S}_\alpha^k(\rho)\leq \mathcal{S}_\alpha^{mk}(\rho)$;
inequalities (\ref{P62}) and (\ref{P65}) hold because  $\bigotimes\limits_{j=1}^m\sigma_j^*\in S_{mk}$ when $\sigma_j^*\in S_k$ and  $\bigotimes\limits_{j=1}^m\widetilde{\sigma}_j^*\in P_{k-1}$ when $\widetilde{\sigma}_j^*\in P_{k-1}$;
Eqs. (\ref{P63})  and  (\ref{P66})  follow from property (A3) of $\alpha$-affinity.
Inequalities (\ref{P64}) and (\ref{P67}) follow from $1-\prod_{i=1}^m x_i \leq \sum_{i=1}^m(1-x_i)$ for $0\leq x_i\leq1$.
Similarly, we can prove that $ \mathfrak{S}_\alpha^k(\rho)$ and $\widetilde{\mathcal{E}}_\alpha^{k}(\rho)$  also satisfy subadditivity.

Since every LOCC operation is a separable operation, the above monotonicity properties also hold under LOCC.

\section{The relationship between multilevel coherence indicators and multipartite correlation indicators}

Let $H$ be a $d$-dimensional Hilbert space. Let $\{|\chi_i\rangle: i=0, 1, \cdots, d-1\}$  be a set that spans $H$ and represents the pure classical states.
The set $C$ of all classical states consists of the convex hull of $\{|\chi_i\rangle\langle\chi_i|\}.$ For a pure state $|\psi\rangle$,  its nonclassical rank is defined as $R_N(|\psi\rangle)=\min\left\{r:|\psi\rangle=\sum\limits_{i=0}^{r-1}c_i|\chi_i\rangle\right\}$;
for a mixed state $\rho$, its nonclassical number is defined as $R_N(\rho)=\min\limits_{\{p_i,|\psi_i\rangle\}}\max\limits_iR_N(|\psi_i\rangle)$, where the minimum is taken over all pure state decompositions $\{p_i,|\psi_i\rangle\}$ of $\rho$.
Several relevant results were established in \cite{RegulaPianiNJP2018}, which are stated as follows:

\emph{Lemma 1.} (Theorem 2 of \cite{RegulaPianiNJP2018})
Let $H$ be a Hilbert space with $\dim(H)=d$, and let
$\{|\chi_i\rangle\}_{i=1}^d$ be a linearly independent set spanning $H$.
Let $H_{\mathrm{anc}}$ denote the Hilbert space of an ancillary system.
Then there exists an isometry
$V: H \rightarrow H \otimes H_{\mathrm{anc}}^{\otimes d}$
such that, for any quantum state $\rho$,
$R_N(\rho)=k
\iff D_E(V\rho V^\dagger)=k+1, 2 \leq k \leq d$,
and
$R_N(\rho)=1 \iff
D_E(V\rho V^\dagger)=1$.

In the proof of this conclusion in  \cite{RegulaPianiNJP2018}, the following notation is used: $|c_i\rangle=|\chi_i\rangle\otimes |\psi_{anc}\rangle\in H\otimes H^{\otimes d}_{anc},$
$|b_i\rangle=|0\rangle^{\otimes i}(\sqrt{\lambda}|0\rangle+\sqrt{1-\lambda}|1\rangle)|0\rangle^{\otimes d-i-1},$
$\{|a_i\rangle: i=0,1,\cdots,d-1\}$ is a linearly independent set,
$U$ is a unitary operator such that $U|c_i\rangle=|a_i\rangle|b_i\rangle.$  It was shown in \cite{RegulaPianiNJP2018} that the
isometry $V$ is defined by the composition of attaching the ancillary state $|\psi_{anc}\rangle$ followed by the action of $U$. Furthermore, it has been shown that
$|\psi'\rangle=V|\psi\rangle=\sum\limits_{i=0}^{d-1}\psi_i|a_i\rangle|0\rangle^{\otimes i}(\sqrt{\lambda}|0\rangle+\sqrt{1-\lambda}|1\rangle)|0\rangle^{\otimes d-i-1}$  for any pure state $|\psi\rangle=\sum\limits_{i=0}^{d-1}\psi_i|\chi_i\rangle\in H$ \cite{RegulaPianiNJP2018}.  Let $S$ be an operator with $S|a_i\rangle=|i\rangle$ and
$L$ be an operator with $L|0\rangle=|0\rangle, L(\sqrt{\lambda}|0\rangle+\sqrt{1-\lambda}|1\rangle)=|1\rangle,$ then $(S\otimes L^{\otimes d})|\psi'\rangle=\sum\limits_{i=0}^{d-1}\psi_i|i\rangle|0\rangle^{\otimes i}|1\rangle|0\rangle^{\otimes d-i-1}.$

Based on the above notation and applying the proof methodology of Lemma 1 (Theorem 2 of \cite{RegulaPianiNJP2018}), we discuss the relationship between $R_N(\rho)$ and $D_S(V\rho V^\dagger)$. By Lemma 1, $D_S(V\rho V^\dagger)=d+1$ if and only if $R_N(\rho)=1$.
The separability depth of $|\psi'\rangle$ is not changed by the local filter $S\otimes L^{\otimes d}$; that is, $D_S(|\psi'\rangle)=D_S(|\widetilde{\psi}'\rangle)$ with $|\widetilde{\psi}'\rangle\propto \sum\limits_{i=0}^{d-1}\psi_i|i\rangle|0\rangle^{\otimes i}|1\rangle|0\rangle^{\otimes d-i-1}.$ This observation also explains why the proof methodology of Lemma 1 (Theorem 2 of \cite{RegulaPianiNJP2018}) can also be applied to separability depth.
If $R_N(|\psi\rangle)=k\geq2,$ then there exist exactly $k$ nonzero coefficients $\psi_i$. Without loss of generality, we may assume that $\psi_0,\psi_1,\cdots,\psi_{k-1}$ are nonzero; that is, $|\psi\rangle=\sum\limits_{i=0}^{k-1}\psi_i|\chi_i\rangle.$
Hence, $|\widetilde{\psi}'\rangle\propto \sum\limits_{i=0}^{k-1}\psi_i|i\rangle|0\rangle^{\otimes i}|1\rangle|0\rangle^{\otimes d-i-1}=(\sum\limits_{i=0}^{k-1}\psi_i|i\rangle|0\rangle^{\otimes i}|1\rangle|0\rangle^{\otimes k-i-1})|0\rangle^{\otimes d-k}$.
The state
$\sum\limits_{i=0}^{k-1}\psi_i|i\rangle|0\rangle^{\otimes i}|1\rangle|0\rangle^{\otimes k-i-1}$
is genuinely multipartite entangled across the first $k+1$ subsystems, and cannot be factorized with respect to any nontrivial partition of these subsystems. Hence, $D_S(|\widetilde{\psi}'\rangle)=d-k+1.$
For $k\geq2$, if $D_S(|\widetilde{\psi}'\rangle)=d-k+1,$ then $R_N(|\psi\rangle)=k.$ (If $R_N(|\psi\rangle)=k'\neq k,$ we obtain $D_S(|\widetilde{\psi}'\rangle)=d-k'+1$ from the above discussion, which leads to a contradiction.) Therefore, for a pure state $|\psi\rangle$, $D_S(|\psi'\rangle)=d-k+1$ if and only if $R_N(|\psi\rangle)=k$ $(2\leq k\leq d).$

There is a one-to-one correspondence between the pure state decompositions $\{p_i,|\psi_i\rangle\}$ of $\rho$ and $\{p_i,|\psi'_i\rangle\}$ of $\rho'=V\rho V^\dagger$ \cite{RegulaPianiNJP2018}. Utilizing the definitions of separability depth and nonclassical number, along with Lemma 1 and the fact that $D_S(|\psi'\rangle)=d-k+1$ if and only if $R_N(|\psi\rangle)=k$ $(2\leq k\leq d),$ we obtain the following conclusion.

\emph{Proposition 1.} $D_S(V\rho V^\dagger)=d-k+1$ if and only if $R_N(\rho)=k$ $(2\leq k\leq d)$, and $D_S(V\rho V^\dagger)=d+1$ if and only if $R_N(\rho)=1$.

When specializing nonclassicality to quantum coherence, the classical pure states are taken to be the orthonormal basis states $\{|i\rangle: i=0,1,\ldots,d-1\}$ of $H$ \cite{RegulaPianiNJP2018}. Let
\[
|\Psi\rangle=|\psi^d\rangle|0\rangle^{\otimes d},
\]
where $|\psi^d\rangle$ is the initial state on $H$ and $|0\rangle^{\otimes d}$ represents $d$ ancillary qubits. Define
\begin{equation}\label{operation}
\begin{aligned}
U=\sum\limits_{i=0}^{d-1}|i\rangle\langle i|\otimes \mathbf{1}^{\otimes i}\otimes\sigma_x\otimes\mathbf{1}^{\otimes d-i-1}.
\end{aligned}
\end{equation}
Then $U$ transforms the state
\[
|\Psi\rangle=\sum\limits_{i=0}^{d-1}c_i|i\rangle|0\rangle^{\otimes d}
\]
into
\[
|\Psi'\rangle=\sum\limits_{i=0}^{d-1}c_i|i\rangle|0\rangle^{\otimes i}|1\rangle|0\rangle^{\otimes d-i-1}.
\]
The following result was provided in \cite{RegulaPianiNJP2018}.

\emph{Lemma 2.} (Theorem 3  of \cite{RegulaPianiNJP2018}) Let $\rho'=U(\rho\otimes|0\rangle\langle0|^{\otimes d})U^\dagger$. Then, for
$2\leq k\leq d$, $R_c(\rho)=k$ if and only if $D_E(\rho')=k+1$.

Similarly, the following conclusion follows from Proposition 1.

\emph{Proposition 2.}  Let $\rho'=U(\rho\otimes|0\rangle\langle0|^{\otimes d})U^\dagger$. Then, for $2\leq k\leq d$,
 $R_c(\rho)=k$ if and only if $D_S(\rho')=d-k+1$.

Based on the above notation, we can state the relationship between multilevel coherence indicators  and multipartite correlation indicators as follows.

\emph{Theorem 3.} Let $\rho$ be a quantum state on a Hilbert space $H$ with $\dim(H)=d$. Define $\rho'=U(\rho\otimes|0\rangle\langle0|^{\otimes d})U^\dagger$, where $U$ is defined by (\ref{operation}). Then, for $3\leq k\leq d+1$, $\mathcal{C}_\alpha^{k}(\rho)\geq \mathcal{S}_\alpha^{d-k+2}(\rho'),$
$\mathfrak{C}_\alpha^{k}(\rho)\geq \mathfrak{S}_\alpha^{d-k+2}(\rho'),$
$\mathcal{C}_\alpha^{k}(\rho)\geq \widetilde{E}_\alpha^{k+1}(\rho'),$ and $\mathfrak{C}_\alpha^{k}(\rho)\geq \widetilde{\mathcal{E}}_\alpha^{k+1}(\rho');$ when $k=2$, $\mathcal{C}_\alpha^{2}(\rho)\geq \mathcal{S}_\alpha^{d+1}(\rho'),$
$\mathfrak{C}_\alpha^{2}(\rho)\geq \mathfrak{S}_\alpha^{d+1}(\rho'),$
$\mathcal{C}_\alpha^{2}(\rho)\geq \widetilde{E}_\alpha^{2}(\rho'),$ and $\mathfrak{C}_\alpha^{2}(\rho)\geq \widetilde{\mathcal{E}}_\alpha^{2}(\rho').$

\emph{Proof.} The proof is similar to that of Theorem 5 of \cite{RegulaPianiNJP2018}, as follows.

For $3\leq k\leq d+1$,
\begin{align}
\mathcal{C}_\alpha^{k}(\rho)=&\min\limits_{\sigma\in I_{k-1}}\Big[1-A_\alpha(\rho,\sigma)\Big]\label{}\nonumber\\
=&\min\limits_{\sigma\in I_{k-1}}\Big[1-A_\alpha(U(\rho\otimes|0\rangle\langle0|^{\otimes d})U^\dagger,U(\sigma\otimes|0\rangle\langle0|^{\otimes d})U^\dagger)\Big]\label{R1}\\
=&\min\limits_{\sigma\in I_{k-1}}\Big[1-A_\alpha(\rho',U(\sigma\otimes|0\rangle\langle0|^{\otimes d})U^\dagger)\Big]\label{}\nonumber\\
\geq&\min\limits_{\sigma'\in S_{d-k+2}}\Big[1-A_\alpha(\rho',\sigma')\Big]\label{R2}\\
=&\mathcal{S}_\alpha^{d-k+2}(\rho').\label{}\nonumber
\end{align}
Here Eq. (\ref{R1}) follows from properties (A1), (A2), and (A3) of  $\alpha$-affinity.

For $\sigma\in I_{k-1}$, $R_c(\sigma)\leq k-1$, thus $D_S(U(\sigma\otimes|0\rangle\langle0|^{\otimes d})U^\dagger)\geq d-k+2$ by Proposition 2. Hence, when $\sigma\in I_{k-1}$, we obtain
$U(\sigma\otimes|0\rangle\langle0|^{\otimes d})U^\dagger\in S_{d-k+2}$. Thus, inequality (\ref{R2}) holds. The remaining inequalities follow analogously from Proposition 2 and Lemma 2.

For $k=2$, when $\sigma\in I_{1}$, one has $R_c(\sigma)=1$ and $U(\sigma\otimes|0\rangle\langle0|^{\otimes d})U^\dagger$  is fully separable, so we obtain $D_S(U(\sigma\otimes|0\rangle\langle0|^{\otimes d})U^\dagger)= d+1$. It follows that
\begin{align}
\mathcal{C}_\alpha^{2}(\rho)=&\min\limits_{\sigma\in I_{1}}\Big[1-A_\alpha(\rho',U(\sigma\otimes|0\rangle\langle0|^{\otimes d})U^\dagger)\Big]\label{}\nonumber\\
\geq&\min\limits_{\sigma'\in S_{d+1}}\Big[1-A_\alpha(\rho',\sigma')\Big]\label{}\nonumber\\
=&\mathcal{S}_\alpha^{d+1}(\rho').\label{}\nonumber
\end{align}
We can similarly prove that the other conclusions also hold for $k=2$.

The relationship between multilevel coherence and  multipartite correlations is summarized in Figure 1.
\begin{figure}[h]
\centering
\includegraphics[width=0.5\textwidth]{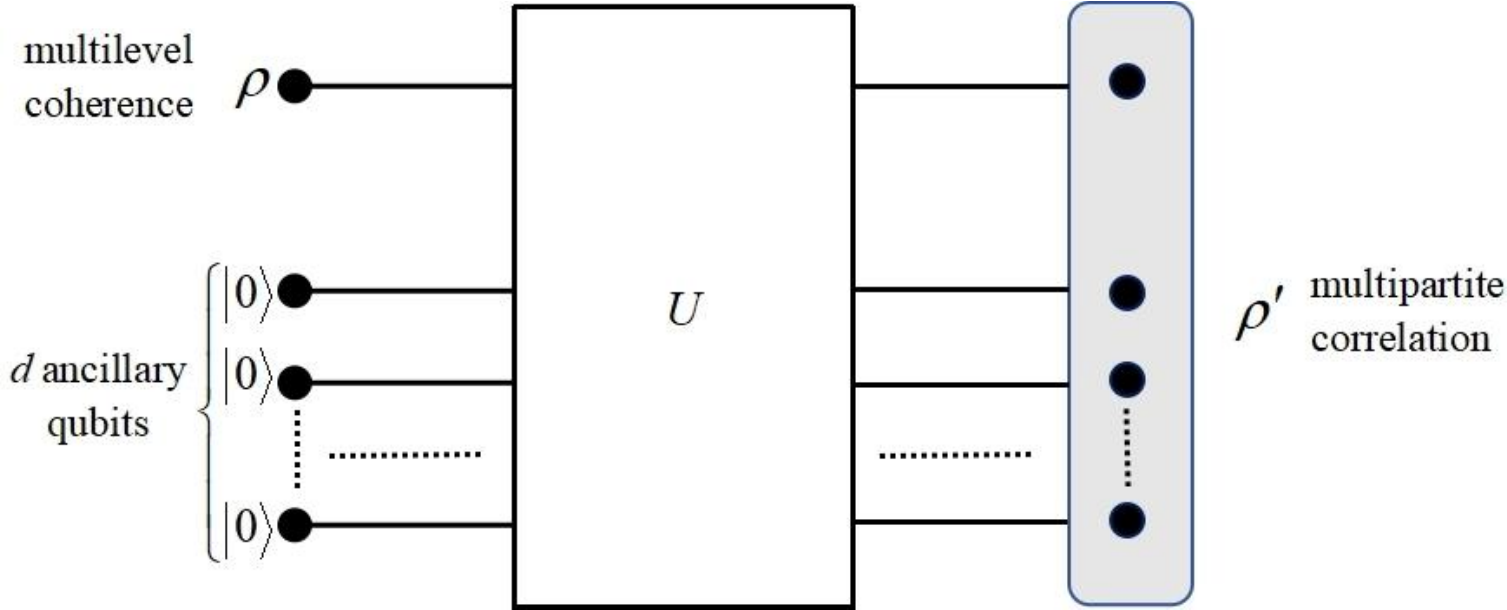}
\caption{ For a quantum state $\rho$ on a Hilbert space $H$, let $U$ be defined by (\ref{operation}).  When $3\leq k\leq d+1$, $\mathcal{C}_\alpha^{k}(\rho)\geq \mathcal{S}_\alpha^{d-k+2}(\rho'),$
$\mathfrak{C}_\alpha^{k}(\rho)\geq \mathfrak{S}_\alpha^{d-k+2}(\rho'),$
$\mathcal{C}_\alpha^{k}(\rho)\geq \widetilde{E}_\alpha^{k+1}(\rho'),$ and $\mathfrak{C}_\alpha^{k}(\rho)\geq \widetilde{\mathcal{E}}_\alpha^{k+1}(\rho')$. Furthermore, it follows that if $\rho$ is a $k$-level coherence-free state, then $\rho'$ is $(d-k+2)$-separable and also $k$-producible.  When $k=2$, $\mathcal{C}_\alpha^{2}(\rho)\geq \mathcal{S}_\alpha^{d+1}(\rho'),$
$\mathfrak{C}_\alpha^{2}(\rho)\geq \mathfrak{S}_\alpha^{d+1}(\rho'),$
$\mathcal{C}_\alpha^{2}(\rho)\geq \widetilde{E}_\alpha^{2}(\rho'),$ and $\mathfrak{C}_\alpha^{2}(\rho)\geq \widetilde{\mathcal{E}}_\alpha^{2}(\rho').$  Moreover, when $k=2$, if $\rho$ is a $2$-level coherence-free state, then $\rho'$ is a fully separable state.}
\end{figure}

\section{Conclusion}

In summary, we have presented two indicators of multilevel coherence and several distinct indicators of multipartite correlations based on $\alpha$-affinity. We have shown that the multilevel coherence indicators $\mathcal{C}_\alpha^k(\rho)$ and $\mathfrak{C}_\alpha^k(\rho)$ are nonnegative and monotonic under $(k-1)$-coherence-preserving operations. In addition, $\mathcal{C}_\alpha^k(\rho)$ is convex, while $\mathfrak{C}_{\alpha}^k(\rho)$ is nonincreasing on average under $(k-1)$-incoherent operations. Both indicators also satisfy the subadditivity property discussed in the main text. Furthermore, we proved that the multipartite correlation indicators $\mathcal{S}_\alpha^k(\rho)$, $\mathfrak{S}_\alpha^k(\rho)$, $\widetilde{E}_\alpha^k(\rho)$, and $\widetilde{\mathcal{E}}_\alpha^k(\rho)$ are nonnegative, invariant under local unitary transformations, monotonic under separable operations (and hence also under LOCC), and subadditive. Among these, $\mathcal{S}_\alpha^k(\rho)$ and $\widetilde{E}_\alpha^k(\rho)$ are also convex, while $\mathfrak{S}_{\alpha}^k(\rho)$ and $\widetilde{\mathcal{E}}_{\alpha}^k(\rho)$ are additionally strongly monotonic under LOCC. We also discussed the relationship between the proposed multilevel coherence indicators and the multipartite correlation indicators, thereby providing a clearer connection between multilevel coherence and multipartite correlations. Given the recent progress in both entanglement distribution \cite{AllenMeyerPRL2017,GourGuoQuantum2018,GuoGourPRA2019,GuoZhangPRA2020,EltschkaSiewertQuantum2018,GuoPRA2024} and coherence distribution \cite{RadhakrishnanParthasarathyPRL2016,MaZhaoPRA2017,XiJPA2018,RadhakrishnanDing2019,BassoMaziero2020},
an interesting direction for future research is to investigate the distribution properties of multilevel coherence. Such investigations may shed new light on the distribution of resources in complex quantum systems.
\begin{center}
{\bf ACKNOWLEDGMENTS}
\end{center}

This work was supported by the National Natural Science Foundation of China under Grant No. 12401604, the Hebei Natural Science Foundation of China under Grant No. A2025403008, and the National Pre-research Funds of Hebei GEO University under Grant No. KY2025YB15.

\appendix

\section{The proof of the properties (A2) and (A3) of $\alpha$-affinity}

\begin{equation*}
\begin{aligned}
A_\alpha(U\rho U^\dagger,U\sigma U^\dagger)=&\textrm{Tr}\Big[(U\rho U^\dagger)^\alpha(U\sigma U^\dagger)^{1-\alpha}\Big]\\
=&\textrm{Tr}\Big[(U\rho^\alpha U^\dagger)(U\sigma^{1-\alpha} U^\dagger)\Big]\\
=&A_\alpha(\rho,\sigma).
\end{aligned}
\end{equation*}
\begin{equation*}
\begin{aligned}
A_\alpha(\rho_1\otimes\rho_2,\sigma_1\otimes\sigma_2)=&\textrm{Tr}\Big[(\rho_1\otimes\rho_2)^\alpha(\sigma_1\otimes\sigma_2)^{1-\alpha}\Big]\\
=&\textrm{Tr}\Big[(\rho_1^\alpha\sigma_1^{1-\alpha})\otimes(\rho_2^\alpha\sigma_2^{1-\alpha})\Big]\\
=&A_\alpha(\rho_1,\sigma_1)A_\alpha(\rho_2,\sigma_2).
\end{aligned}
\end{equation*}

\section{The proof of inequality (\ref{PropositionB5})}

\emph{Lemma B1.} (Theorem B of \cite{Tian2014}) If $a_m>0$, $b_m>0$, $p>0$, $q<0$, and $u=\max\left\{1,\frac1p+\frac1q\right\}$, then
$$\sum\limits_{m=1}^na_mb_m\geq n^{1-u}\Big(\sum\limits_{m=1}^na_m^p\Big)^{\frac{1}{p}}\Big(\sum\limits_{m=1}^nb_m^q\Big)^{\frac{1}{q}}.$$

\emph{Proposition B1.} If $x_m>0, q_m>0, t>1$, and $\sum\limits_{m=1}^nq_m=1$, then
\begin{equation}\label{PropositionB1}
\begin{aligned}
\sum\limits_{m=1}^nx_m^tq_m^{1-t}\geq \Big(\sum\limits_{m=1}^nx_m\Big)^t.
\end{aligned}
\end{equation}

\emph{Proof.}  Let $a_m=x_m^t$, $b_m=q_m^{1-t}$, $p=\frac{1}{t}$, and $q=\frac{1}{1-t}$, then inequality (\ref{PropositionB1}) follows from Lemma B1.

\emph{Proposition B2.} For a CPTP map with Kraus operators $\{K_i\}$, let $p_i=\textrm{Tr}(K_i\rho K_i^\dagger), q_i=\textrm{Tr}(K_i\sigma K_i^\dagger), \rho_i=\dfrac{K_i\rho K_i^\dagger}{p_i}$, and $\sigma_i=\dfrac{K_i\sigma K_i^\dagger}{q_i}$. Then
\begin{equation*}\label{PropositionB2}
\begin{aligned}
\sum\limits_ip_i\Big[A_\alpha(\rho_i,\sigma_i)\Big]^{\frac{1}{\alpha}}\geq\Big[\sum\limits_iA_\alpha(p_i\rho_i,q_i\sigma_i)\Big]^{\frac{1}{\alpha}}.
\end{aligned}
\end{equation*}

\emph{Proof.} Using the definition of $\alpha$-affinity and Proposition B1, we obtain
\begin{align}
\sum\limits_ip_i\Big[A_\alpha(\rho_i,\sigma_i)\Big]^{\frac{1}{\alpha}}
=&\sum\limits_iq_i^{1-\frac{1}{\alpha}}\left\{\textrm{Tr}\Big[(K_i\rho K_i^\dagger)^\alpha(K_i\sigma K_i^\dagger)^{1-\alpha}\Big]\right\}^{\frac{1}{\alpha}}\label{}\nonumber\\
\geq&\left\{\sum\limits_i\textrm{Tr}\Big[(K_i\rho K_i^\dagger)^\alpha(K_i\sigma K_i^\dagger)^{1-\alpha}\Big]\right\}^{\frac{1}{\alpha}}\nonumber\\
=&\Big[\sum\limits_iA_\alpha(p_i\rho_i,q_i\sigma_i)\Big]^{\frac{1}{\alpha}}.\label{}\nonumber
\end{align}

\emph{Proposition B3.} Let $\{K_i\}$ be the Kraus operators of a CPTP map on the original system, and let $\{|i_a\rangle\}$ be an orthonormal basis of the auxiliary system. Then
\begin{equation*}\label{PropositionB3}
\begin{aligned}
A_\alpha(\sum\limits_iK_i\rho K_i^\dagger\otimes|i_a\rangle\langle i_a|,\sum\limits_iK_i\sigma K_i^\dagger\otimes|i_a\rangle\langle i_a|)=\sum\limits_iA_\alpha(K_i\rho K_i^\dagger,K_i\sigma K_i^\dagger).
\end{aligned}
\end{equation*}

\emph{Proof.} Let $P$ be a unitary permutation operator that exchanges the original system and the auxiliary system. Then
\begin{align}
&A_\alpha(\sum\limits_iK_i\rho K_i^\dagger\otimes|i_a\rangle\langle i_a|,\sum\limits_iK_i\sigma K_i^\dagger\otimes|i_a\rangle\langle i_a|)\label{}\nonumber\\
=&\textrm{Tr}\Big[(\sum\limits_iK_i\rho K_i^\dagger\otimes|i_a\rangle\langle i_a|)^\alpha(\sum\limits_iK_i\sigma K_i^\dagger\otimes|i_a\rangle\langle i_a|)^{1-\alpha}\Big]\label{}\nonumber\\
=&\textrm{Tr}\Big[(P\sum\limits_iK_i\rho K_i^\dagger\otimes|i_a\rangle\langle i_a|P^\dagger)^\alpha(P\sum\limits_iK_i\sigma K_i^\dagger\otimes|i_a\rangle\langle i_a|P^\dagger)^{1-\alpha}\Big]\label{PropositionB31}\\
=&\textrm{Tr}\Big[\sum\limits_i|i_a\rangle\langle i_a|\otimes (K_i\rho K_i^\dagger)^\alpha (K_i\sigma K_i^\dagger)^{1-\alpha}\Big]\label{PropositionB32}\\
=&\sum\limits_i\textrm{Tr}\Big[( K_i\rho K_i^\dagger)^\alpha (K_i\sigma K_i^\dagger)^{1-\alpha}\Big]\label{}\nonumber\\
=&\sum\limits_iA_\alpha(K_i\rho K_i^\dagger,K_i\sigma K_i^\dagger).\label{}\nonumber
\end{align}
Equation (\ref{PropositionB31}) follows from the unitary invariance of $\alpha$-affinity. Equation (\ref{PropositionB32}) follows from the definition of the permutation operator $P$.

\emph{Proposition B4.} For a CPTP map with Kraus operators $\{K_i\}$, the following relation holds,
\begin{equation*}\label{PropositionB4}
\begin{aligned}
A_\alpha(\rho,\sigma)\leq\sum\limits_iA_\alpha(K_i\rho K_i^\dagger,K_i\sigma K_i^\dagger).
\end{aligned}
\end{equation*}

\emph{Proof.} Let $\{|i_a\rangle\}$ be an orthonormal basis of the auxiliary system. For any channel $\Lambda$ with Kraus operators $\{K_i\}$, there exists a unitary operator $U$ such that the following equality holds \cite{YuPRA2017}:
\begin{equation}\label{PropositionB41}
\begin{aligned}
K_i\rho K_i^\dagger\otimes |i_a\rangle\langle i_a|
=(\textbf{1}\otimes|i_a\rangle\langle i_a|)
\,U(\rho\otimes|0_a\rangle\langle 0_a|)U^\dagger\,
(\textbf{1}\otimes|i_a\rangle\langle i_a|),
\end{aligned}
\end{equation}
where $U$ acts on the composite system consisting of the original system and the auxiliary system. Thus we obtain
\begin{align}
A_\alpha(\rho,\sigma)=&A_\alpha(U(\rho\otimes|0_a\rangle\langle 0_a|)U^\dagger,U(\sigma\otimes|0_a\rangle\langle 0_a|)U^\dagger)\label{PropositionB42}\\
\leq&A_\alpha\Big[\varepsilon(U(\rho\otimes|0_a\rangle\langle 0_a|)U^\dagger),\varepsilon(U(\sigma\otimes|0_a\rangle\langle 0_a|)U^\dagger)\Big]\label{PropositionB43}\\
=&A_\alpha\Big[\sum\limits_i\textbf{1}\otimes|i_a\rangle\langle i_a|U(\rho\otimes|0_a\rangle\langle 0_a|)U^\dagger\textbf{1}\otimes|i_a\rangle\langle i_a|,\sum\limits_i\textbf{1}\otimes|i_a\rangle\langle i_a|U(\sigma\otimes|0_a\rangle\langle 0_a|)U^\dagger\textbf{1}\otimes|i_a\rangle\langle i_a|\Big]\label{PropositionB44}\\
=&A_\alpha(\sum\limits_iK_i\rho K_i^\dagger\otimes|i_a\rangle\langle i_a|,\sum\limits_iK_i\sigma K_i^\dagger\otimes|i_a\rangle\langle i_a|)\label{PropositionB45}\\
=&\sum\limits_iA_\alpha(K_i\rho K_i^\dagger,K_i\sigma K_i^\dagger).\label{PropositionB46}
\end{align}
Equation (\ref{PropositionB42}) follows from properties (A1), (A2), and (A3) of $\alpha$-affinity. Inequality (\ref{PropositionB43}) follows from property (A5). Equation (\ref{PropositionB44}) follows from the CPTP map defined by the Kraus operators $\{\mathbf{1}\otimes |i_a\rangle\langle i_a|\}$. Equations (\ref{PropositionB45}) and (\ref{PropositionB46}) follow from Eq. (\ref{PropositionB41}) and Proposition B3, respectively.

\emph{Proposition B5.} For a CPTP map with Kraus operators $\{K_i\}$, let $p_i=\textrm{Tr}(K_i\rho K_i^\dagger), q_i=\textrm{Tr}(K_i\sigma K_i^\dagger), \rho_i=\dfrac{K_i\rho K_i^\dagger}{p_i}$, and $\sigma_i=\dfrac{K_i\sigma K_i^\dagger}{q_i}$. Then
\begin{equation*}
\begin{aligned}
\sum\limits_ip_i\left\{1-\Big[A_\alpha(\rho_i,\sigma_i)\Big]^{\frac{1}{\alpha}}\right\}\leq1-\Big[A_\alpha(\rho,\sigma)\Big]^{\frac{1}{\alpha}}.
\end{aligned}
\end{equation*}

\emph{Proof.} By Proposition B2 and Proposition B4, we obtain
\begin{align}
\sum\limits_ip_i\left\{1-\Big[A_\alpha(\rho_i,\sigma_i)\Big]^{\frac{1}{\alpha}}\right\}
=&1-\sum\limits_ip_i\Big[A_\alpha(\rho_i,\sigma_i)\Big]^{\frac{1}{\alpha}}\label{}\nonumber\\
\leq&1-\Big[\sum\limits_iA_\alpha(p_i\rho_i,q_i\sigma_i)\Big]^{\frac{1}{\alpha}}\label{}\nonumber\\
\leq&1-\Big[A_\alpha(\rho,\sigma)\Big]^{\frac{1}{\alpha}}.\label{}\nonumber
\end{align}

\end{document}